\documentclass[12pt,a4paper]{article}

\usepackage{amssymb, amsfonts, amsthm, graphics, graphicx}

\usepackage{epsfig, euscript}

\begin{document}

\title{Prequantum Classical Statistical Field Theory: Simulation of Probabilities of Photon 
Detection with the Aid of Classical Brownian Motion}

\author{Andrei Khrennikov\\International Center for Mathematical Modeling in Physics,\\ Engineering, Economics, and Cognitive Science\\
Linnaeus University, S-35195, V\"axj\"o-Kalmar, Sweden\\
 Artem Bazhanov\\
Moscow Institute of Electronic Technology\\ 124498, Zelenograd, Moscow, Russia Federation}

\maketitle

\begin{abstract} In this paper we present results of numerical simulation based on 
Prequantum Classical Statistical Field Theory (PCSFT), a model with hidden variables of 
the field-type reproducing probabilistic predictions of quantum mechanics (QM). PCSFT is combined 
with measurement theory based on detectors of the threshold type. The latter describes discrete
events corresponding to the continuous fields model, PCSFT. Numerical modeling demonstrated that the
classical Brownian motion (the Wiener process valued in complex Hilbert space) producing clicks
when approaching the detection threshold gives probabilities of detection predcited by the formalism 
QM (as well as PCSFT). This numerical result is important, since  the transition from PCSFT to 
the threshold detection has a complex mathematical structure (in the framework of 
classical random processes) and it was modeled only approximately. We also perform numerical simulation 
for the PCSFT-value of the coefficient of second order coherence. Our result matches well with the
 prediction of quantum theory. Thus, opposite to semiclassical theory, PCSFT cannot be rejected as 
a consequence of measurements of $g^{(2)}(0).$ Finally, we analyze the output of the recent experiment \cite{Z1} 
performed in NIST questioning the validity of some predictions of PCSFT.  
\end{abstract}

keywords: Prequantum classical field theory; foundations of quantum mechanics; numerical simulation of discrete events; 
Brownian motion in complex Hilbert space; threshold detectors; coefficient of second order coherence

\section{Introduction}

At the very beginning Einstein's idea that the electromagnetic field can be quantized even in vacuum, i.e., 
that its quanta play the fundamental role not only in  the process of the energy exchange with matter\footnote{ {\small In 1910 Einstein wrote:
``What we understand by the theory of ``light quanta" may be formulated in the following fashion: a
radiation of frequency $\nu$ can be emitted or absorbed only in a well defined quantum
of magnitude $h\nu.$ The theoreticians have not yet even come to an agreement in
regard to the following question: Can the light quanta be accounted for entirely by a
characteristic of the emitting or absorbing substance, or should the electromagnetic
radiation itself be assigned, besides a wave structure, such that the energy of the
radiation itself is already divided in definite quanta? I believe that I have proven
that this latter view should be adopted,"} \cite{6}, p. 207. }, 
was not commonly accepted. In particular, Planck strongly opposed it.
At the beginning (until 1920th)  even Bohr was not happy
with the invention of light quanta. In particular, the Bohr-Kramers-Slater theory\cite{1}
was an attempt to describe the interaction of matter and electromagnetic radiation
without using the notion of photon. However, very soon the majority in the quantum community 
accepted  ``photon's existence''. Nevertheless, two of ``fathers of QM'', namely   
Lande \cite{2,3}  (in particular, this
name is associated with Lande $g$-factor and the first explanation for the anomalous
Zeeman effect) and Lamb \cite{4} (e.g., Lamb shift), see also Lamb and Scully \cite{5},
did not accept quantization of the electromagnetic field, they were supporters of 
the {\it semi-classical model} of QM: the matter is quantized, but the radiation not; the quantum-like 
features of radiation are exhibited only in the process of interaction with quantized matter.  
Here discrete events which by the orthodox interpretation of QM are identified with photons are just 
clicks of detectors.´We also remark that the idea of photon is completely foreign to 
stochastic electrodynamics, see, e.g., De la Pena and Cetto,  Nieuwenhuizen  \cite{v}--\cite{v2}. 

Recently a new approach to classical field modeling of quantum phenomena was presented  \cite{19}--\cite{DT4} in the 
framework of so-called {\it Prequantum Classical Statistical Field Theory} (PCSFT), a model with hidden variables of 
the field-type reproducing probabilistic predictions of quantum mechanics. The main distinguishing feature of PCSFT
is that this is a model of the purely wave type, i.e., there are no particles. PCSFT  can be considered as a comeback 
to the early Schr\"odinger views: attempts to identify the wave function with a classical physical field. Later 
Schr\"odinger gave up and accepted the probabilistic interpretation of the wave function proposed by Born.

For Schr\"odinger, the main problem was the impossibility of the physical field interpretation of the wave function of a 
compound system. 
He pointed out that 
the wave function of a single electron is defined on physical space which mathematically represented as ${\bf R}^3.$ However, the wave 
function of a pair of electrons is defined on the configuration space ${\bf R}^6$ and cannot be considered as a physical field. Pauli 
formulated this problem as the dilemma: {\it Either physical fields on unphysical spaces or virtual (probabilistic) fields on physical space.} 
This interpretation
problem for the wave function was solved in PCSFT. Here the wave function is not directly identified with a physical field, but can be used 
to describe such a field mathematically. The difference in the interpretations of the wave function in QM and PCSFT can be shortly described as 
follows: in QM the wave function directly generates probabilities (therefore it is so convenient to consider it as a field of probabilities) and in 
PCSFT it also generates probabilities, but indirectly, through generation of a classical random field.\footnote{So, operationally 
it is easier to apply the formalism of QM than the formalism of PCSFT. In the latter determination of probabilities contains 
additional steps which are mathematically nontrivial. However, the formal operational interpretation of QM induces numerous mysteries and
exoticisms which are absent in PCSFT.}   
The correspondence between quantum states and, so to say, subquantum random fields is based on the relation (\ref{EERR7}).    

PCSFT was completed with measurement theory based on detectors of the threshold type \cite{DT1}--\cite{DT4}. The latter describes {\it discrete
events corresponding to the continuous fields model}, PCSFT. Numerical modeling presented in this paper demonstrated that the
classical Brownian motion (the Wiener process valued in complex Hilbert space) producing clicks
when approaching the detection threshold gives probabilities of detection predicted by the formalism 
QM (as well as PCSFT). This numerical result is important, since  the transition from PCSFT to 
the threshold detection has a complex mathematical structure (in the framework of 
classical random processes) and it was modeled only approximately.

The main idea behind PCSFT \cite{19} is that the quantum density operator $\rho$   can be obtained from theory of classical random fields 
as normalization by the trace of the covariance operator
$B$ of a subquantum field $\phi(s)$:
\begin{equation}
\label{EERR7}
\rho =\frac{B}{\rm{Tr} B}.
\end{equation}
This field can be considered as  classical field representation of a quantum system. Thus the basic 
notion of QM, the notion of the quantum state, lost its fundamental value. In PCSFT it is an emergent notion, it is reduced to the well 
known notion of classical probability theory, density operator.

We remark that in PCSFT all random fields have zero average.     
Even for such fields, in general the covariance $B$ does not determine the random field  $\phi(s)$ uniquely.
However, if one restricts the class of subquantum fields to the Gaussian fields, then the correspondence
between fields and covariance operators is one-to-one. However, even for Gaussian subquantum fields 
the correspondence  (\ref{EERR7}) is not one-to-one, since it involves the normalization constant 
$\sigma^2= \rm{Tr} B.$ Its probabilistic meaning is the dispersion of the subquantum random field;
its physical meaning is the power of the field. (Thus in PCSFT it is possible to define the power of
a ``photon''.)

The next step \cite{DT1}--\cite{DT4} was creation of measurement theory corresponding to PCSFT and describing transition from
continuous subquantum fields to discrete events, ``clicks of detectors.''  We proceed under the assumption that all detections are 
detections of the threshold type.   In our model a detector is an operational entity $D$ such that 
 a click is produced when the energy of the random field interacting with $D$ approaches
the {\it detection threshold} ${\cal E}_d$. It is easy to present physical models of classical devices interacting with
the electromagnetic field in this way. We can speculate that even quantum detectors work in
this way. We proceed  by considering a black box  $D$ which produces a ``click'' when the energy level of  the random field 
 $\phi(s)$ becomes larger than the detection threshold ${\cal E}_d.$ 
It was  shown \cite{DT1}--\cite{DT4} that (surprisingly) discrete events,
clicks, produced by such operational entities can reproduce 
the quantum probabilities of detection.

We also perform numerical simulation 
for the PCSFT-value of the {\it coefficient of second order coherence} \cite{10}. Our result matches well with the
 prediction of quantum theory  \cite{10}. Thus, opposite to semiclassical theory, PCSFT cannot be rejected as 
a consequence of measurements of $g^{(2)}(0)$ \cite{Grangier}-\cite{Beck} (also known as experiments 
confirming the existing of photons).  Finally, we analyze the output of the recent experiment \cite{Z1}
performed in NIST questioning the validity of some predictions of PCSFT \cite{DT2}--\cite{DT4}.

Modeling of discrete events has the justification function for the theoretical model of PCSFT. It also provides 
a computer model of real processes, cf.  De Raedt and  Michielsen et al  \cite{Raedt}--\cite{Raedt3} 
(in these papers extended modeling of quantum phenomena in terms of discrete 
events was performed, from the double slit experiment to Bell's inequality).

In this paper we restrict considerations to classical modeling of
``finite-dimensional QM'', i.e., the state space of subquantum fields
is finite dimensional. Therefore we shall call such random fields (in accordance
with the classical probabilistic terminology) {\it stochastic processes.}

\section{Clicks as the results of interaction of threshold detectors with classical stochastic processes}

In this section we briefly repeat the detection scheme presented in \cite{DT1}--\cite{DT4} and derivation of detection probabilities 
corresponding to this scheme as well as their matching with probabilities given by QM, under the assumption that 
the correspondence between PCSFT and QM is based on (\ref{EERR7}). We repeat that our mathematical considerations are 
done at the ``physical level of rigorousness'', see Remark 1.  

\subsection{Single channel detection scheme}
\label{SDSD77}

We consider a threshold type detector $D$ with the threshold ${\cal E}_d.$ It interacts with a
stochastic process $\phi(s; \omega),$ where $s$ is time and $\omega$ is a chance parameter describing randomness. 
For a moment, we consider a ${\bf C}$-valued stochastic process, a complex stochastic process, signal. 

The energy of the signal $\phi(s; \omega)$ is given by ${\cal E}(s; \omega) =  \vert \phi(s; \omega) \vert^2$  (hence, the stochastic process
has the physical dimension $\sim
\sqrt{\rm{energy}})$. A threshold detector clicks at the first
moment of time $\delta= \delta(\omega),$  when the energy of the signal ${\cal E}$ exceeds the threshold:
\begin{equation}
\label{E3}
{\cal E} (\delta(\omega), \omega) \geq {\cal E}_d.
\end{equation}

After this event the detector's state updates and $D$ is ready to interact with the next pulse. 
(In reality there is also 
a ``dead time'' period when $D$ cannot interact with a newcommen pulse. In this paper, in particular, in 
coming numerical simulation  we ignore this ``technicality''.)
It is assumed that there is a source ${\cal S}$
 of pulses and each pulse in the process of interaction with $D$  is transformed into a signal. Realizations of this 
signal (inside $D)$ are represented by the stochastic process $\phi(s; \omega)$ under consideration. Thus $\phi(s; \omega)$ is not the original signal, say  
$\phi_0(s; \omega),$ emitted by ${\cal S},$ but the result of its interaction with $D.$ It is natural to assume that the basic probabilistic
characteristics of $\phi(s; \omega)$ are determined by the corresponding characteristics of $\phi_0(s; \omega).$ In particular, we shall 
proceed under the assumption:

\medskip

{\bf In=Out(Prob)} {\it The process emitted by ${\cal S},$ the input for $D,$ and the process $\phi(s; \omega)$ have the same average 
(in fact, zero average)  and that their 
covariances coincide.}

\medskip   

This, although completely classical, picture of detection matches well Bohr's views on quantum measurements \cite{BR0}--\cite{P4}.
 He emphasized many times 
that the whole experimental arrangement has to be taken into account. And we stress the role of the detector changing the form 
of the signal. 

Our argument leading to taking into account the transformer role of $D$ is not straightforward and it was not presented in so much 
details in the previous publications on PCSFT and its measurement model \cite{DT2}, \cite{DT4}. Therefore now we spend some time to clarify 
treatment of $D$ as not only a device for registration of signal's overcoming the threshold  ${\cal E}_d,$ but also as transformer 
of signal's stochastic features. Since we assume {\bf In=Out(Prob)} and since, as we shall see, the detection probabilities are expressed
solely in terms of the covariance operator, it seems that the model can be simplified by identifying the signal $\phi(s; \omega)$ producing 
clicks on the basis of the decision rule (\ref{E3}) with the source output $\phi_0(s; \omega).$ However, as we shall see, the temporal-spatial 
form of the trajectories of stochastic process $\phi(s; \omega)$ do not match with the ``pulse form'' of signals emitted 
by  ${\cal S}.$ Therefore to match with the pulse, wave-packet, picture of emitted signals and at the same time to match with 
quantum detection probabilities,  we have to endow  $D$ not only with the registration feature, see (\ref{E3}),
but even with a transformation feature,   $\phi_0 \to \phi.$ 
 
\medskip

In the mathematical model the detection moment is defined as the {\it first hitting
time}
\begin{equation}
\label{E4}
\delta(\omega) = \inf\{s \geq 0: {\cal E} (\delta(\omega), \omega) \geq {\cal E}_d\}.
\end{equation}

Up to now, we have proceeded with arbitrary stochastic signals. 
To get detection probabilities matching with the formalism of QM, we have 
to select stochastic processes of a special class. For a moment, we cannot 
describe mathematically the class of processes generating quantum detection 
probabilities for the correspondence  (\ref{EERR7}) between theory of classical 
stochastic processes and QM, by taking into account {\bf In=Out(Prob)}. However, we found a few simple examples of Gaussian 
processes which produce such a matching \cite{DT2}, \cite{DT4}. The simplest one is the Brownian 
motion. Thus, after arriving into a threshold type
detector the classical stochastic process $\phi(s; \omega)$
behaves inside this detector as the Brownian motion in the space of (complex) fields.

Thus the process $\phi(s, \omega)$ is the Wiener process: the Gaussian process having zero average, 
 $E\phi(s, \omega)=0$, and 
covariance
$
E\phi(s_1, \omega) \overline{\phi(s_2, \omega)}
 = \min(s_1, s_2) \sigma^2; 
$ 
we can find average of its energy
$E{\cal E} (s, \omega)=\sigma^2 s.$
We find that the coefficient
$\sigma^2 = \frac{E{\cal E} (s, \omega)}{S}$ has the physical dimension of {\it power.}

We are interested in average of the moments of the ${\cal E}_d$-threshold detection for
the energy of the Brownian motion. Since moments of detection are defined formally
as hitting times, we can apply theory of hitting times for the Wiener process, see
e.g. Shyryaev:
$\bar{\delta}\equiv E \delta = \frac{{\cal E}_d}{\sigma^2}$ 
or
\begin{equation}
\label{E8}
\frac{1}{\bar{\delta}}=\frac{\sigma^2}{{\cal E}_d} \;.
\end{equation}
Hence, during a long period of time $T$ such a detector clicks $N_\sigma$-times, where
\begin{equation}
\label{E9}
N_\sigma\approx \frac{T}{\bar{\delta}}=\frac{\sigma^2 T}{{\cal E}_d} \;.
\end{equation}

\medskip

{\bf Remark 1.} This formula determining the number of clicks in the channel by using the average detection time
is approximate; in general, this is sufficiently rough approximation. Therefore coming numerical simulation, section \ref{NSIM},
plays the important role in justification of the presented threshold detection scheme.     

\subsection{Complex Wiener process as a class of stochastic processes}
\label{ONE}

The notion of the complex Wiener process is more complicated than it was formally presented in the previous section. 
In fact, each $b=\sigma^2$ determines a class of  real  Wiener processes valued in $\mathbf R^2$ and, hence, a class of stochastic processes valued
in $\mathbf C.$ One may say that the terminology used in the previous section was misleading and it might be better from the very beginning 
to speak about a class of stochastic processes. However, there is a point in the aforementioned terminology. All processes from the class
of stochastic processes determined by a ``complex Wiener process" determine the same probability distribution of clicks of detectors. Hence,
they can be considered as equivalent from the operational viewpoint. (We remark that they are not equivalent from the viewpoint of theory of stochastic 
processes.) This operational approach is closer to quantum formalism (if the latter is also interpreted as an operational formalism), see section 
\ref{HH77}.

Now we discuss the multi-process structure of the complex Wiener process. We start with the trivial remark 
that in general the easiest way to construct a complex valued process $\xi(t)$ is to use the combination of two real processes,
$\xi_1(s), \xi_2(s),$ namely, set $\xi(s)= \xi_1(s)+i \xi_2(s).$ If the processes $\xi_1(s), \xi_2(s)$ are independent, 
 then the dispersion of $\xi(t)$ equals to the sum of dispersions of $\xi_1(t)$ and  $\xi_2(t).$

Now consider an arbitrary factorization of $b=\sigma^2$ in the form $b= c \bar c,$ where $c \in \mathbf C, c=k_1, + i k_2, k_j \in  \mathbf R.$
Consider now the standard $\mathbf R^2$-valued Wiener process \cite{G} $w(s)= (w_1(s), w_2(s)),$ where $w_1(s), w_2(s)$ are independent one dimensional 
Wiener processes: $Ew_j^2(s)=1, j=1,2,$ and $Ew_1(s) w_2(s)=0.$ We now scale it by setting $w^\prime(s) = \frac{1}{\sqrt{2}} w(s),$ i.e., now 
 $E (w^\prime_j(s))^2=1/2, j=1,2.$ This scaling is needed, since a  complex  process is combined of two real components and 
its total dispersion is the sum of  components' dispersions. We set $W(s)= w^\prime_1(s) + i w^\prime_2(s).$  We call this stochastic 
process the {\it standard complex Wiener process.} (This is the concrete complex valued stochastic process determined by a pair of 
real Wiener processes.) Its dispersion $E \vert W(s) \vert^2= E     (w^\prime_1(s))^2 + (w^\prime_2(s))^2 =1.$ Now to obtain a $b$-process, we simply scale 
$W(s): \phi(s)=c W(s).$ Then $E \vert \phi(s) \vert^2= b.$ It is interesting to consider the real representation 
of this process:
\begin{equation}
\label{88}
\phi(s)= [k_1 w^\prime_1(s) - k_2 w^\prime_2(s)] + i [k_2 w^\prime_1(s) + k_1 w^\prime_2(s)].
\end{equation}
In fact, the above consideration represented the algorithm of construction of stochastic processes which we used for numerical simulation.
However, in the applications to the quantum measurement problem we need vector valued  processes, see section  \ref{LIo}
for analogous considerations 
in the multi-dimensional case. 

Thus the object which we call complex Wiener process is a class of real valued two dimensional Wiener processes. 
They are not arbitrary; they have very special structure, see (\ref{88}). This structure is a consequence of symplectic 
invariance of real processes representing complex processes with the aid of scaling, see section  \ref{LIo} and the paper 
\cite{19}. We shall discuss possible physical interpretation of this non-uniqueness in section \ref{LIo}.

\subsection{Multi-channel detection scheme}

Consider now a stochastic process $\phi(s; \omega)$ valued in the $m$-dimensional complex Hilbert
space $H.$

 Let $(e_j)$ be an orthonormal basis in $H.$ The vector-valued stochastic process 
$\phi(s, \omega)$ can be expanded with respect to this basis
\begin{equation}
\label{NN7}
\phi(s; \omega) = \sum_j \phi_j(s; \omega) e_j, 
\end{equation}
where $\phi_j(s; \omega)= \langle \phi(s; \omega)\vert e_j\rangle.$
In accordance with our detection scheme, section \ref{SDSD77}, the process 
$\phi(s; \omega)$ is mathematical representation of the vector of physical signals 
constructed in the following way. The source emits the random signal  $\phi_0(s; \omega) \in H.$
(For example, in the case of consideration of only polarization degrees of freedom $H$ is the two dimensional 
Hilbert space; here we neglect the spatial degrees of freedom.)  This process is split, e.g., by polarization beam splitter
into components propagating in disjoint channels,  $j = 1, 2,...,m.$ Here 
$
\phi_0(s; \omega) = \sum_j \phi_{0j}(s, \omega) e_j. 
$
We now assume that there is a threshold detector in
each channel, $D_1, ..., D_m.$ Then each component $\phi_{0j}(s; \omega)$ 
through interaction with $D_j$ is transformed into a signal mathematically represented by the stochastic process 
$\phi_{j}(s; \omega).$ The vector with these components can be represented as the $H$-valued stochastic 
process  $\phi(s; \omega),$ see (\ref{NN7}). And we proceed under the vector-form of the assumption {\bf In=Out(Prob)}, section 
\ref{SDSD77}.

We again remark that the scheme world be essentially simpler and it would lead to the same result, if we omit the detector-transformation step
and operate directly with process emitted by the source as belonging to the class generating quantum detection probabilities (e.g., the 
Wiener processes), cf. with discussion
in  section \ref{HH77}.
However, we prefer the above more complex scheme, since it is convenient to use for description of 
propagation in space-time (which is, in fact, absent in this paper, but possible \cite{}) 
one type of stochastic processes and in detectors as another type. 

We also assume that all detectors have the same threshold
${\cal E}_d > 0.$

To each detector we apply the above detection scheme.
The crucial point is that the detectors work totally independently 
from each other. An event in one detector, a click, has no influence 
on the physical processes in other detectors. If in some detector $D_j$  a click is generated, then 
for this detector the experimental trial is finished and a new trial is started. What is about other detectors? 
As was pointed out, they work without ``paying any attention'' to such an event in  $D_j.$
Another detector $D_i, i\not=j,$ continues its interaction with signal's component   
$\phi_i(s; \omega)$ until signal's energy will  overcome its detection threshold.        
Thus events in detectors, clicks, are not sharply coupled
events in the source, namely, emission of pulses. 
It can happen that the clicks are produced simultaneously in a few detectors. In this case we register all them. 
In numerical simulation we cannot work with the notion  
``simultaneously''; there will be always assumed the presence of a nontrivial {\it time window} used to identify clicks 
as ``simultaneously occurring.''

\subsection{Wiener process valued in complex Hilbert space}
\label{HH77}

Suppose now that   $\phi(s, \omega) $  is 
the Wiener process valued in $H$. 
This process is determined by the covariance
operator $B: H \to H.$ Any covariance operator 
is Hermitian, positive, and trace-class  and vice versa. The complex Wiener process is characterized
by Hermitian covariance operator. We have, for $y\in H,$
$
E \langle y, \phi(s, \omega )\rangle =0, 
$ 
and, for $y_j\in H, j=1,2,$ 
$E \langle y_1, \phi(s_1, \omega )\rangle \langle \phi(s_2, \omega ), y_2 \rangle =\min(s_1,s_2) \langle B y_1, y_2 \rangle.
$ 
The latter is the covariance function of the stochastic process; in the operator form:
$
B(s_1, s_2) = \min(s_1,s_2) B.
$
We note that the dispersion of the $H$-valued Wiener process (at the instant of time $s)$ is given by 
$\Sigma^2(s)= E \Vert \phi(s, \omega) \Vert^2= s \rm{Tr} B.
$
The quantity ${\cal E}(s, \omega)= \Vert \phi(s, \omega) \Vert^2$ 
is the total energy of the Brownian motion signal at the instant of time $s.$ Hence, the quantity
\begin{equation}
\label{E11}
\Sigma^2\equiv \frac{\Sigma^2(s)}{s} = \rm{Tr} B
\end{equation} 
is the average power of this random signal.  

We also remark that by normalization of 
the covariance function for the fixed $s$ by the dispersion 
we obtain the operator,
\begin{equation}
\label{E12}
\rho= B(s,s)/\Sigma^2(s)= B/ \rm{Tr} B,
\end{equation} 
which formally has all properties of 
the {\it density operator} used in 
quantum theory to represent quantum states.
Its matrix elements have the form $\rho_{ij} = b_{ij}/\Sigma^2.$ These
are dimensionless quantities. 

The relation (\ref{E12}) is basic for PCSFT : 

\medskip

Each classical random process generates a quantum state (in general
mixed) which is given by the normalized covariance operator of the process. 

\medskip

Consider components $\phi_j(s, \omega)$ of the vector valued signal $\phi(s, \omega).$
Then $$ 
E \phi_i(s, \omega)\overline{\phi_j(s, \omega)}= \min(s_1,s_2) \langle B e_i, e_j \rangle= b_{ij}.$$
In particular, 
$\sigma_j^2(s) \equiv  E {\cal E}_j(s,\omega)\equiv E \vert  \phi_j(s, \omega)\vert^2= s b_{jj}.
$
This is the average energy of the $j$th component at the instant of time $s.$ We also consider 
the average powers of components
$\sigma_j^2 \equiv \frac{\sigma_j^2(s)}{s}= b_{jj}.$
We remark that the average power of the total signal is equal to the sum of  the average powers of its components.
$\Sigma^2= \sum_{j} \sigma_j^2.$

\subsection{Detection probabilities, the multi-channel case}

Consider now a run of experiment of the duration $T.$ The average number of clicks
for the $j$th detector can be approximately expressed as
\begin{equation}
\label{E15}
N_j \equiv N_{\sigma_j}\approx \frac{\sigma_j^2 T}{{\cal E}_d}.
\end{equation}
The total number of clicks is 
(again approximately) given by 
$N= \sum_j N_j \approx \frac{\Sigma_j^2 T}{{\cal E}_d}.$
Hence, for the detector $D_j,$  the probability of detection can be expressed as
\begin{equation}
\label{E16}
P_j \approx \frac{N_j}{N} \approx \frac{\sigma_j^2}{\Sigma_j^2} = \rho_{jj},
\end{equation}
however, see Remark 1.

This is, in fact, {\it the Born's rule for the quantum state $\rho$  and the projection operator
$\hat{C}_j = \vert e_j\rangle \langle e_j\vert$  on the vector $e_j.$}  For the detector $D_j,$  the probability of detection can
be expressed as
\begin{equation}
\label{E17}
P_j = \rm{Tr} \rho \; \hat{C}_j. 
\end{equation}

\subsection{Wiener process in complex Hilbert space as a class of vector-valued stochastic processes}
\label{LIo}

Now we briefly repeat considerations of section \ref{ONE} in the multi-dimensional case. Let $B$ be a positively defined Hermitian 
matrix. Consider one of its representations in the form:
\begin{equation}
\label{88a}
B= C C^*,
\end{equation}
where $C$ is an arbitrary complex matrix. This $C$ can be represented as $C=K_1 + i K_2,$ where $K_j, j=1,2,$ are real matrices.
In the same way we represent  $B$ as $B=B_1+i B_2.$ Thus  
the equation (\ref{88a}) can be written as the system of equations:
\begin{equation}
\label{88a}
B_1= K_1 K_2^*+ K_2 K_2^*, \; B_1= K_2 K_1^* - K_1 K_2^*. 
\end{equation}
Consider now the standard complex $m$  dimensional Wiener process \cite{G} ${\bf W}(s) =(W_1(s),..., W_m(s)),$ each its coordinate is one dimensional 
standard complex Wiener process, see section \ref{ONE}, and the they are independent.   This process has the unit covariation operator:
\begin{equation}
\label{88b8}
E[\langle {\bf W}(s;  \omega)\vert u \rangle \overline{\langle {\bf W}(s;  \omega)\vert v\rangle}= \langle u\vert v\rangle,
\end{equation}
where $u,v$ are two arbitrary vectors.

To obtain the $B$-process, we simply scale 
the standard complex Wiener process ${\bf W}(s):$
\begin{equation}
\label{88b}
 \phi^C(s)= C {\bf W}(s).
\end{equation}
(In section \ref{NSIM} we shall use this representation to generate 
a complex Wiener $B$-process; one of its versions.)
It is easy to check that the process $\phi^C(s)$ 
really has the covariance operator $B:$ 
$$
E [\phi^C_j(s; \omega) \overline{\phi^C_i(s; \omega)}] = E[ \langle \phi^C(s; \omega)\vert e_j\rangle \overline{\langle 
\phi^C(s; \omega)\vert e_i\rangle}]=
$$
$$
E[\langle {\bf W}(s;  \omega)\vert C^*e_j\rangle 
\overline{\langle {\bf W}(s;  \omega)\vert C^*e_i\rangle }] = 
\langle C^*e_j \vert C^*e_i\rangle=
\langle B e_j \vert e_i\rangle =b_{ji}, 
$$
see (\ref{88b8}).

Consider now decomposition of $\phi^C(s)$ into real Wiener processes. For this, we first decompose the standard complex Wiener process 
${\bf W}(s) = {\bf w}^\prime_1(s) + i {\bf w}^\prime_1(s),$ where ${\bf w}^\prime_j(s), j=1,2,$ are $1/\sqrt{2}$-scalings of two 
standard (independent) real Wiener processes in the $m$-dimensional space. 
Then
\begin{equation}
\label{88t}
\phi^C(s)= [K_1 {\bf w}^\prime_1(s) - K_2 {\bf w}^\prime_2(s)] + i [K_2 {\bf w}^\prime_1(s) +  K_1 {\bf w}^\prime_2(s)].
\end{equation}

We remark that all processes of the form (\ref{88a}) generate the same average of energy in each of detectors:
\begin{equation}
\label{88ta}
E \vert \phi^C_i(s; \omega)\vert^2 = b_{ii}.
\end{equation}
And our model predicts that the statistics of clicks in detectors depends only on the (relative) average energy delivered to each detector
(for a random signal in a detector which can be represented by a complex Wiener process). Therefore it is not surprising that 
a variety of processes (\ref{88b}) with (\ref{88a}) produce the same probability distribution of clicks in the detectors.
One cannot distinguish these processes by using the detectors. From the PCSFT-viewpoint, the quantum formalism is an operational
formalism. Here a density matrix encodes  a variety of subquantum processes. Can one hope to design experiments to approach such processes?
This is an open problem.

\section{Numerical modeling for detection probabilities}
\label{NSIM}

As was pointed out in Remark 1, in our mathematical modeling of detection probabilities for the classical Wiener process interacting with 
the threshold detector we used rough approximations and through these approximate calculations we approached matching with the 
quantum formula for probabilities of the results of measurement, Born's rule, see  (\ref{E17}). We state again that the prequantum$\to$ quantum correspondence
by itself is exact, see (\ref{EERR7}). However, we were not able to solve exactly the problem of theory of classical stochastic processes, the problem 
of expression of probabilities for the threshold detection through elements of the covariance matrix of the process. Now we would like to complete
the approximate theoretical picture by numerical modeling. We produced a plenty of graphical data corresponding to such simulation. 

We start with  
the data corresponding to dichotomous observables, expressed in terms of the two channel model.
Consider the two dimensional complex Hilbert space. There is given a quantum observables, 
say $\hat{A},$ let $(e_1, e_2)$ be the basis consisting of its 
eigenvectors. Consider the Wiener process in this space with the covariance operator $B,$
the subquantum process induced as the result of interaction of ``quantum systems'' with detectors
measuring $\hat{A}.$ In the eigen-basis, the covariance operator is represented by the corresponding matrix:  
$
B=\left(\begin{array}{cc}
b_{11} & b_{12}\\
b_{21} & b_{22}
\end{array}\right),
$  
where $b_{11}, b_{22}$ are real numbers and $\bar{b}_{12}= b_{21}.$ To consider the general situation, we do not assume that 
$\rm{Tr} B=1,$ i.e., the prequantum$\to$ quantum map (\ref{EERR7}) is not identity, but contains nontrivial normalization by 
$b_{11}  + b_{22}.$ 

We now present an example of numerical simulation. We use the scheme which was presented in section \ref{LIo}. Here 
with the covariance matrix has the form
\begin{equation}
\label{77}
B=\left(\begin{array}{cc}
10 & 5+2i\\
5-2i & 9
\end{array}\right)=SS^*,
\end{equation}
and we selected 
\begin{equation}
\label{77s}
S=\left(\begin{array}{cc}
1 & 3i\\
2-2i & i
\end{array}\right),
\end{equation}
Here $P_1=10/(10+9)=10/19 \approx 0. 526, P_2= 9/(10+9)=9/19 \approx 0. 474.$  The relative frequencies
$\nu_1=N_1/(N_1+N_2), \nu_2=N_2/(N_1+N_2),$ where $N_j; j=1,2,$ are the numbers of clicks in corresponding detectors.  
The results of simulation showed that the relative frequencies approach the predicted 
probabilities, although they fluctuate at the initial segment of series of trials. 


We also present the results of the multi-dimensional simulation. We consider the covariance matrix
\begin{equation}
\label{77a}
B=\left(\begin{array}{cccc}
14, & 4-2i &-2-5i & 7-4i \\
4+2i & 12 & -7-i & 2\\
-2+5i &-7+i & 8 & 1+4i \\
7+4i & 2 & 1-4i & 6
\end{array}\right) = SS^*,
\end{equation}
and we selected
\begin{equation}
\label{77d}
S= \left(\begin{array}{cccc}
2-2i & i & 1 & 2\\
1 & 3i &1 &-1 \\
i &-2i & i &1+i \\
2 &0 &1 & 1
\end{array}\right),
\end{equation}
Here our model predicts the probabilities of detection: 
$P_1=14/40=0.35,      P_2= 12/40=0.3,         P_3= 8/40=0.2,
P_4= 6/40=0.15.$
The results of simulation showed that the relative frequencies approach the predicted 
probabilities, although they fluctuate at the initial segment of series of trials. 


\section{Numerical modeling for the coefficient of second order coherence}
\label{TT}

\begin{figure}[htpb]
\centering{}
\includegraphics[scale=1]{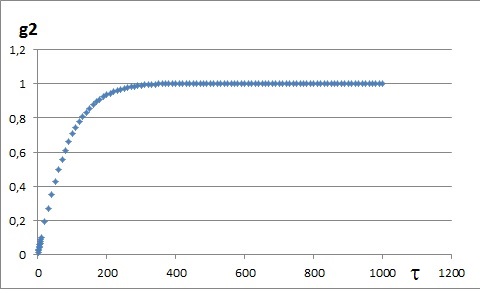} 
\caption{The graph of $g^{(2)}$ depending on the coincidence time window $\tau$ (for the $B$-Wiener process with 
the covariance matrix given by (\ref{77}) 
and the threshold ${\cal E}_d=\rm{Tr} B/20.$}
\label{FIG3} 
\end{figure}

We remark that quantum theory predicts  \cite{10} that, for single photon states, the coefficient of second order coherence $g^{(2)}(0) =0.$ At the same 
time for semiclassical models $g^{(2)}(0) \geq 1.$ Therefore measurements of $g^{(2)}(0)$ played the crucial role in distinguishing 
the quantum and semiclassical models of micro-phenomena \cite{10}-\cite{Beck}. Such experiments were also crucial to confirm experimentally the 
``existence of photon'' \cite{Grangier}, \cite{Grangier1}. Therefore it is important to find the magnitude of  the coefficient of second order coherence in the framework of PCSFT 
(completed with the threshold model of measurement). It is difficult to find  $g^{(2)}(0)$ analytically by using theory of classical stochastic 
processes; in \cite{DT2}--\cite{DT4} there was found its estimate from above showing that at least for the special inter-relation between the 
detection threshold and signal's energy this coefficient is less than 1. In this paper we use numerical simulation for this purpose. 
The coefficient of second order coherence is defined as 
\begin{equation}
\label{88d}
 g^{(2)}(0)= \frac{P_{12}}{P_1 P_2},
\end{equation} 
where $P_j$ are the probabilities of detection in the channels $j=1,2$ and $P_{12}$ is the probability of the coincidence of clicks in the two detectors. 
In previous sections, to find $P_j,$ we used the normalization by the sum of clicks in the two channels. This approach did not take into account 
the coincidence detections. Now we consider the normalization  which is proper for calculation of $ g^{(2)}(0)$ (and more generally  
$g^{(2)}(s), s \geq 0).$ By considering the sum $N=N_1+N_2$ we take the number of coincidence clicks $N_{12}$ twice. Therefore, for the proper
normalization, we have to use $N= N_1+N_2 - N_{12}.$ Hence, in this section we set 
\begin{equation}
\label{81}
 P_j=\frac{N_j}{N_1+N_2 - N_{12}}, P_{12} = \frac{N_{12}}{N_1+N_2 - N_{12}}, 
\end{equation}
Numerical simulations shows that for our model the quantity $N_{12}/(N_1+N_2)$ is small. Therefore the values of probabilities 
calculated in the previous 
sections change only slightly, so the graphs at Fig. \ref{FIG1}, Fig. \ref{FIG2} are indistinguishable from the graphs corresponding to  
the frequency-probabilities given by (\ref{81}).   We obtain the following expression for $g^{(2)}(0):$
\begin{equation}
\label{81}
g^{(2)}(0)= \frac{N_{12}(N_1+N_2 - N_{12})}{N_1 N_2}.
\end{equation}
This formula can be applied for numerical simulation and experiment.
However, there is one pitfall which is typically not 
taken into account in the discussions on calculation of 
$g^{(2)}(0)$ (see, however, \cite{Beck}). Both in numerical simulation and experiment 
we cannot proceed with continuous time and the corresponding notion 
of coincidence ``at the same instance of time.'' A coincidence time 
window $\tau$ has to be used. Therefore the quantities depend on this 
time window: $N_1= N_1(\tau), N_2= N_2(\tau), N_{12}= N_{12}(\tau)$ and, hence,
$g^{(2)}(0; \tau).$   The graph for $g^{(2)}(0; \tau)$ as the function of the time window
$\tau$ is presented at Fig. \ref{FIG3}. We see that for small time widows the coefficient of second order
coherence is strictly less than 1 and it goes to zero for $\tau \to 0.$ This is in the complete 
agreement with the prediction of QM. Hence, opposite to semiclassical optics, the PCSFT cannot be rejected 
as the result of experimental measuring of the coefficient of second order coherence.

Besides the graph at Fig. \ref{FIG3}, it is also useful to present the numerical values of $g^{(2)}(0; \tau)$ for various
values of the time window $\tau=1,2,...9,10, 20,..., 50:$\\ 
$g^{(2)}= 0,0112465; 0,0216078; 0,0314473;  0,039609; 0,0484313;  0,0574435;\\
0,0676744; 
0,0779637;
0,0860406; 
0,0972437; 
0,191344; 
0,271691; 
0,350341; 
0,428744.$

In classical and quantum optics not only $g^{(2)}(0),$ but also $g^{(2)}(s), s \geq 0,$ plays an important role \cite{10}. 
In fact, our function $g^{(2)}(0; \tau), \tau \geq 0,$ can be considered as a good approximation for  
$g^{(2)}(\tau), \tau \geq 0.$ And we see again that the graph at Fig. \ref{Fig3} matches well with the corresponding graph 
from quantum theory \cite{10}. 

Finally, we remark that recently a group of experimenters from NIST performed the experimental test \cite{Z1} challenging 
the predictions of PCSFT about the dependence of the coefficient  $g^{(2)}(0)$ on the inter-relation of the detection 
threshold and the signal energy \cite{DT2}--\cite{DT4}. They did not find dependence of the form predicted in \cite{DT2}--\cite{DT4}. 
There are a few objections to consider this experiment as a negative test for PCSFT (with the measurement model based on the threshold detection). 

The first  remark is that in \cite{DT2}--\cite{DT4} we were able only to get an estimate from above and the upper bound 
depends on the aforementioned parameters. This, of course, does not imply that   $g^{(2)}(0)$ by itself depends on them in this way;
it might be that this is just a bad estimate. Unfortunately, in the present paper we were not able to model dependence of $g^{(2)}(0)$ on the 
detection threshold and energy to approach the regime considered in \cite{DT2}--\cite{DT4}, i.e., for example, 
very high levels of the detection threshold
compare with $\rm{Tr} B.$ Here the Brownian motion exceeds the detection threshold not so often and the available computational resources
were not sufficient to perform such a numerical simulation (one has to use really powerful computer to collect good statistics).

 Another objection to the interpretation of the results of the NIST-experiment \cite{Z1} as confronting with PCSFT is that the experimenters did not use 
the real single photon source; they worked with so-called heralded photons. The latter technique is standard in the recent experiments on the
calculation of the coefficient of second order coherence. We remark that even the original experiment of Grangier \cite{Grangier}, 
\cite{Grangier1} was done with heralded
photons. Typically the coefficient calculated with the aid of heralded photons is not distinguished from the ``genuine coefficient of second order 
coherence'', i.e., calculated with the aid of ``really single photon sources''. However, here one has to careful, since these are, in fact, 
two different quantities, although both reflect the same quantum feature of light.
   
\section{Concluding remarks and further studies}

In this paper we confirmed with the aid of numerical simulation the theoretical predictions of the threshold detection model for PCSFT
which were obtained in \cite{DT2}--\cite{DT4}. As was pointed out, in these works calculations were done at the physical level of rigorousness,
quantum mechanical predictions were approached only asymptotically and without estimation of the degree of approximation. Therefore 
numerical simulation presented here is very important for justification of the predictions of  \cite{DT2}--\cite{DT4}.

\section*{Acknowledgments}

One of the authors (A. Khrennikov) would like to thank experimenters from NIST, S. Polyakov and A. Migdall, and from Institute for Quantum Optics and Quantum 
Information (QOQI),  S. Ramelow, R. Ursin, and B. Wittmann, for critical discussions on various possibilities
to falsify PCSFT experimentally and A. Plotnitsky for as well critical discussions on the interpretation issues related to PCSFT,
the authors also thank I. Basieva for help with numerical simulation. This study was partially supported through visiting professor fellowship 
of A. Khrennikov to QOQI and this author would like to thank A. Zeilinger for hospitality and fruitful discussions on quantum foundations;  A. Khrennikov 
also was supported by the grant of the Faculty of Technology of Linnaeus University ``Mathematical 
Modeling of Complex Hierarchic Systems''. The visit of A. Bazhanov to Linnaeus University was based on the grant on International Students Exchange of 
Moscow Institute of Electronic Technology.

\end{document}